\documentstyle[12pt]{article}
\begin{document}
\baselineskip = 20pt

\begin{flushright}
{CERN-TH/95-259}
\end{flushright}
\vspace*{5mm}

\begin{center}

{DILATON GRAVITY\\ WITH A NON-MINIMALLY COUPLED\\ SCALAR FIELD}

\vspace*{1cm}

{M. ALVES$^{\star}$\\
\vspace{0.3cm}
Theory Division - CERN\\
CH-1211 GENEVE 23 \\
Switzerland}\\
\vspace*{2mm}
{ABSTRACT}
\end{center}
\vspace*{5mm}

\noindent
We discuss the two-dimensional dilaton gravity with a scalar field as
the source matter. The coupling between the gravity and the scalar,
massless, field is presented in an unusual form. We work out two
examples of these couplings, and solutions with black-hole behaviour are
discussed and compared with those found in the literature.

\bigskip
\vfill
\noindent PACS: 04.60.+n; 11.17.+y; 97.60.Lf
\par
\bigskip
\bigskip
\noindent CERN-TH/95-259

\noindent October 1995
\par
\bigskip
\bigskip
\noindent $^\star${e-mail: ALVES@VXCERN.CERN.CH}

\noindent On leave from \,
 Instituto de Fisica, Universidade Federal do Rio de Janeiro, Brazil.

\pagebreak
\noindent
{\bf 1 INTRODUCTION}
\par It is widely recognized that two-dimensional models of gravity can give
us a better understanding of the gravitational quantum effects. These
models, derived from a string motivated effective action [1] or from
some low-dimensional version of the Einstein equations [2], have a rich
structure in spite of their relative simplicity. Gravitational collapse,
black holes and quantum effects are examples of subjects whose description
is rather complicated in four-dimensional gravity while their lower-dimensional
version turns out to be more treatable, sometimes completely solved.

This is the case of the seminal work of Callan, Giddings, Harvey and
Strominger (CGHS)[1], where the Hawking effect is analysed
semiclassically. Black-hole solutions are found and used to obtain an
expression for the Hawking radiation.
Some improvements have been done [3] to circumvent difficulties that
arise from
its quantum version, but leaving the initial purpose unchanged.

In the CGHS model, one starts with the four-dimensional Einstein-Hilbert
action in the spherically symmetric metric. Then, the assumption of
dependence in two variables  of all the fields is done.
With a suitable form of the metric, it is possible to integrate
the angular dependence. The resulting equations can be derived from
a two-dimensional action, the dilaton being the relic of the
integrated coordinates.
Then this action is modified, rendering  a simpler one, which is
called the two-dimensional dilaton gravity.

With this modified action, black-hole solutions are found to be formed from
non-singular initial conditions, namely a source of scalar matter,
coupled minimally with the two-dimensional gravity sector (the terminology
will be clarified later). It is shown that there is also a linear vacuum
region,
which imposes conditions to the expression for the Hawking radiation.

In this model, the scalar matter is coupled only with the
two-dimensional gravity sector,
without interaction with the dilaton field. Since this
field is part of the formerly four-dimensional world and, as pointed out in
[1],
some results obtained are model- dependent, it could be interesting to analyse
some consequences in considering the coupling between the scalar matter and
the dilaton field, as originally derived from the dimensional reduction.
This is the main purpose of
the present work. It is organized as follows: we begin with the construction
of the 2d dilaton gravity, but in an unusual way, and then
modifications
are introduced with the respective solutions. Discussion and final remarks
are in the conclusion.

\bigskip
\bigskip
\bigskip
\noindent
{\bf 2  THE 2d DILATON GRAVITY AND THE DIMENSIONAL REDUCTION}
\par
Intending to compare results, we present in this section
the CGHS model. However, this will be done in a slightly different way,
closer to the dimensional reduction, giving more motivations for the
modifications we will work out.

The starting point is the Einstein action on $\bar D$ dimensions coupled with a
dilaton field $\bar\phi$ [4]

\begin{equation}
S=\int d^{\bar D}x \sqrt{-{\bar g}}\, e^{-\bar\phi}
\biggl\{ \bar g^{\mu\nu}\partial_\mu\bar\phi
\partial_\nu\bar\phi +\bar R \biggr\},
\end{equation}

\noindent where $\bar R$ is the curvature scalar in $\bar D$ dimensions
and $ \bar D=D+d $.
When $\bar D = 10$, this is the bosonic part of the heterotic string
effective action in the critical dimension [4]. The antisymmetric part
was omitted,  since our interest is in the 2d case.

The theory is considered in a spacetime $ M \times K $, where $M$ and
 $K$ are $D$ and $d$ dimensional spacetime, respectively. If the fields
are independent of the $d$ coordinates, the $\bar D$ metric can be
decomposed as [4]

\begin{equation}
 \bar g_{\bar\mu\bar\nu} = \left(
\begin{array}{cc}

g_{\mu\nu} + A_{\mu}^{\gamma} A_{\nu\gamma} & A_{\mu\beta} \\

A_{\nu\alpha} & G_{\alpha\beta}
\end{array}
\right).
\end{equation}

\noindent The resulting dimensionally reduced action is

\begin{equation}
S=\int d^Dx \sqrt{-g}\, e^{-\phi}\, \biggl\{ R + g^{\mu\nu}\partial_\mu\phi
\partial_\nu\phi
+ {1\over 4} g^{\mu\nu}\partial_\mu G_{\alpha\beta}
\partial_\nu G^{\alpha\beta} - {1\over 4} g^{\mu\rho}g^{\nu\lambda}
G_{\alpha\beta}F_{\mu\nu}^{\alpha}F_{\rho\lambda}^{\beta} \biggr\}
\end{equation}

\noindent with $\phi=\bar\phi - {1\over 2}\,log \, (det\, G_{\alpha\beta})$
being the
shifted dilaton and $F_{\mu\nu}^{\alpha} = \partial_{\mu}A_{\nu}^{\alpha} -
\partial_{\nu}A_{\mu}^{\alpha}$.

Now we make the  assumptions to reach the action we are interested in,
namely taking the simple case where  the original dilaton field
$\phi$ is zero and writing the metric in the most general spherically
symmetric form:

\begin{equation}
ds^{2} = g_{\mu\nu}dx^{\mu}dx^{\nu}
+ {1\over{\lambda^{2}}}e^{-2\phi}d^{2}\Omega.
\end{equation}

So, with $D=d=2$ and for $A_{\mu\gamma}=0$
we have the 2d dimensionally reduced action

\begin{equation}
S={1\over{2\pi}}\int d^{2}x \sqrt{-g}\, e^{-2\phi} \biggl\{ R +
2(\nabla\phi)^{2} +  2\lambda^{2}e^{2\phi} \biggr\}.
\end{equation}

\noindent It is worth to mentioning that this is the simplest example of
the dimensional reduction giving an action with $O(d,d)_{d=2}$ symmetry.
If one starts with a scalar field, action (1) changes to

\begin{equation}
S= \int d^{\bar D}x \sqrt{-\bar g}\, e^{-\phi} \biggl\{ \bar g^{\mu\nu}
\partial_\mu \bar\phi \partial_\nu \bar\phi + \bar R
- {1\over 2}\bar g^{\mu\nu}\partial_\mu \bar f \partial_\nu \bar f \biggr\}
\end{equation}

and we have, with the same assumptions:

\begin{equation}
S= \int d^{2}x \sqrt{-g}\, e^{-2\phi} \biggl\{ R + 2(\nabla \phi )^2 +
2\lambda^2 e^{2\phi} - {1\over 2}(\nabla f)^2 \biggr\}
\end{equation}

\noindent for the reduced version. Note that in (7),
the scalar matter couples with both
the 2d-gravity sector and the dilatonic field.  We will consider this
coupling as the first modification to the CGHS action.
The second one concerns the
semiclassical quantization and some explanation is needed.

The semiclassical quantization via functional method requires integrating
over fields (the scalar one in this particular case). On the other hand, we
are interested in theories that have the full quantized version free of
anomalies. This paradigm can be worked out via the BRST analysis of the
theory by using the Fujikawa's technique [6]. In this framework, the quantized
theory is anomaly-free if the functional measure is BRST-invariant. It follows
 that this
invariance requires some redefinition of the field variables, the so-called
gravitational dressing, and we must consider it as the field variable of the
model that we are studying. It is instructive to show how the gravitational
dressing works in a very simple example, the 2d conformal scalar field,
minimally coupled with gravity.

 The  first-order
quantum-corrected
version for the 2d scalar field gives a traceless vacuum expected
value (VEV) for the  energy
momentum tensor

\begin{equation}
\langle T^{\mu}_{\mu}\rangle = g^{\mu\nu} \langle T_{\mu\nu}\rangle = 0
\end{equation}

\noindent but non covariantly conserved

\begin{equation}
\nabla^\nu\langle T_{\mu\nu}\rangle =
\langle\partial_\mu f \nabla^\nu\partial_\nu f\rangle
\end{equation}

\noindent since (9) is an ill-defined quantity.
However, most of the quantities of interest arise from
non-classical effects (e.g. the Hawking radiation) it
could be useful to consider the
theory within this scope.  At the same time, we hope that the
model (in the
semiclassical version) agrees
with some basic principles, one of them being the (semiclassical) EMT
covariantly conserved [5], unlike the one given by (9).

This undesired result can be avoided by a redefinition of the
field variable [7]:

\begin{equation}
f \rightarrow  \tilde f = g^{({1\over 4})} f.
\end{equation}

\noindent It is straightforward to see that, considering $\tilde f$
as the variable,

\begin{equation}
\langle \tilde T^\mu_\mu \rangle =
\langle f \nabla^\mu\partial_\mu f \rangle = {1\over 24\pi}R
\end{equation}

\noindent and

\begin{equation}
\nabla^\nu \langle \tilde T_{\mu\nu} \rangle = 0,
\end{equation}

\noindent to renormalized vacuum expected values.
The classical action for the new field variable is obtained from the old one,
using the definition (10):

\begin{equation}
S[\tilde f]= \int d^2x (\nabla \tilde f)^2.
\end{equation}

In the four-dimensional case, the redefinition of the field variable is
not sufficient to give us the conservation of the VEV of the energy-momentum
tensor. Instead of this, we must impose (12) as a condition to be satisfied.
Actually, it is used to eliminate ambiguities from the definition of some
numerical factors in the expression of the renormalized VEV of the
four-dimensional energy-momentum tensor [5].

{}From the definition (10) and the coupling given by (7), these follows a
modified action to the dilaton gravity

\begin{equation}
S= {1\over {2\pi}} \int d^2x \, e^{-2\phi}\biggl\{ \sqrt{-g}
[ R + 4(\nabla\phi)^2 + 4\lambda^2 ] - {1\over 2}(\nabla\tilde f)^2
\biggr\},
\end{equation}

\noindent where we have done the same modifications, regarding the
numerical factors and the coupling between the dilaton and the constant
$\lambda$, as those in the CGHS model. Since there are couplings other
than the original dilaton gravity action, we call the theory given by
(14) the non-minimally coupled one.

\bigskip
\bigskip
\bigskip
\noindent
{\bf 3  BLACK-HOLE SOLUTIONS FROM SCALAR MATTER}
\par In this section, we show that there are black-hole-type solutions
arising from a pulse of scalar matter, as in the CGHS model. However,
due to the new couplings with both the dilaton field and the
two-dimensional gravity sector, these solutions have new features
with respect to those from the usual dilatonic gravity model.

The equations of
motion, derived from (14), are

\begin{equation}
e^{-2\phi}[\nabla_\mu\nabla_\nu\phi + g_{\mu\nu}((\nabla\phi)^2 - \nabla^2\phi
-\lambda^2) - {1\over 2}\nabla_\mu f\nabla_\nu f ] = 0,
\end{equation}

\begin{equation}
e^{-2\phi} [ R + 4\lambda^2 + 4\nabla^2\phi - 4(\nabla\phi)^2 ] = 0
\end{equation}

\noindent and

\begin{equation}
\nabla^2 f = 2\nabla_\mu f\nabla^\mu\phi \, ,
\end{equation}

\noindent
for the dilaton, the metric, and the scalar field, respectively. The first two
equations imply

\begin{equation}
e^{-2\phi}[ R + 2\nabla^2\phi ]=0.
\end{equation}

It is useful to work with light-cone coordinates

\begin{equation}
x^{\pm} = x^0\pm x^1
\end{equation}

\noindent and choose the conformal gauge $g_{\mu\nu} = e^{2\rho}\eta_{\mu\nu}$
that, in
the light-cone coordinates, gives

\begin{equation}
g_{+-} = -{1\over 2} e^{2\rho}\, , \,\,\,\, g_{++} = g_{--} = 0.
\end{equation}

\noindent Now, using (17), the $(+-)$ component of the equation for the
dilaton (15) becomes

\begin{equation}
\partial_+\partial_-\phi - 2\partial_+\phi\partial_-\phi =
{1\over 2}\lambda^2 e^{2\rho} - {1\over 2}\partial_+f\partial_-f
\end{equation}

\noindent or

\begin{equation}
\partial_+\partial_-(e^{-2\phi}) =
 -\lambda^2 + \partial_+f\partial_-f e^{-2\phi}
\end{equation}

\noindent and eq.(18)

\begin{equation}
\partial_+\partial_-\phi = \partial_+\partial_-\rho.
\end{equation}

\noindent We can use the residual gauge freedom [1]
to make

\begin{equation}
\rho=\phi.
\end{equation}

\noindent This relation, besides rendering calculations easier, has some
implications on the original four-dimensional metric (4). We will comment
on this point later.

The equations of motion for the gauge-fixed components $g_{++}$ and
$g_{--}$ must be imposed as constraints:

\begin{equation}
\nabla_+\nabla_+\phi = {1\over 2} \nabla_+f\nabla_+f \,\,\,\,\,\,
and \,\,\,\,\,\, \nabla_-\nabla_-\phi = {1\over 2} \nabla_-f\nabla_-f.
\end{equation}

\noindent Using (24) and  $ \Gamma_{++}^+ = 2\partial_+\rho $ and
$ \Gamma_{--}^- = 2\partial_-\rho $ \, these relations can be written as

\begin{equation}
\partial_+^2\phi - 2(\partial_+\phi)^2 = {1\over 2}(\partial_+f)^2
\end{equation}

\noindent and

\begin{equation}
\partial_-^2\phi - 2(\partial_-\phi)^2 = {1\over 2}(\partial_-f)^2.
\end{equation}

Let us consider, as a source of matter, a shock-wave travelling in the
$x^-$ direction. As in [1], this will be described by the
stress tensor

\begin{equation}
{1\over 2}\partial_+f\partial_+f = a \delta (x^+ - x_{0}^+),
\end{equation}

\noindent where $a$ is its magnitude.

Defining $h(x^+,x^-)=e^{-2\phi}$ eqs. (26) and (27) can be written  as

\begin{equation}
\partial_+^2h = -a \delta (x^+-x_{0}^+)h
\end{equation}

\noindent and

\begin{equation}
\partial_-^2h = 0.
\end{equation}

We note that there is a difference between eq.(28) and the corresponding
one from the dilaton gravity. This difference is due to the coupling we have
introduced in the action (14). The solution of this new equation has the
following general form:

\begin{equation}
h(x^+,x^-) = - a h(x^+_{0},x^-)x^+ + A(x^-)x^+ + K,
\end{equation}

where $A(x^-)$ is a function of only  $x^-$ and $K$, a constant. The quantity
$h(x^+_{0},x^-)$ can be evaluated taking (31) at $x^+= x^+_{0}$,  giving

\begin{equation}
h(x^+_{0},x^-) = {K\over {1+ax^+_{0}}} + {x^+_{0} \over {1+ax^+_{0}}} A(x^-).
\end{equation}

Using the eq.(22), the final expression for the solution is

\begin{equation}
h(x^+,x^-) = -\lambda^2 x^+x^{-\prime} + {M\over\lambda},
\end{equation}

\noindent with $x^{-\prime} = x^- + a{c_1\over\lambda^2}$
and $c_1 = {M\over {\lambda (1+ax^+_{0})}}$ .

Comparing this expression with the one from the dilaton gravity, we see
that (33) corresponds to a black-hole solution with mass given by $M=K\lambda$
and the coordinate $x^-$ shifted to

\begin{equation}
x^{-\prime} = x^- + a{M\over{\lambda^3 (1+ax^+_{0})}}.
\end{equation}

\noindent When $a=0$, which means no source, we have a situation called
the linear vacuum solution, also present in the CGHS model, in the
same situation.

Another situation can be easily analysed, namely when there are
two matter sources,
one of them being like (28) and the other given by

\begin{equation}
\partial_-f\partial_-f = b\delta(x^--x^-_{0})
\end{equation}

The only modification with respect to the previous case is in the constraint
equation
(27), for the $g_{--}$ component, written as

\begin{equation}
\partial_+^2h(x^+,x^-) = -b\delta(x^--x^-_{0}).
\end{equation}

\noindent It is straightforward to show that the solution is

\begin{equation}
h(x^+,x^-) = -\lambda^2 x^+x^- + K,
\end{equation}

\noindent that is the eternal black-hole solution, with $K={M\over\lambda}$,
found in the CGHS model,
when there are no sources. This result is not completely unexpected: as
seen at the first case, the non-minimal coupling introduces solutions with the
same behaviour in both  $x^+<x^+_{0}$ and $x^+>x^+_{0}$ regions.
When the source (35) is
present, the regions are separated by this front wave that turns to be
unchanged, and no horizon is presented. We can see this directly in eq.(37).

\bigskip
\bigskip
\bigskip
\noindent
{\bf 4  CONCLUSIONS AND FINAL REMARKS}
\par In the last section, we have analysed the solutions generated from
the non-minimal coupling between the scalar matter and both the
dilaton field and the 2-d gravity sector. This coupling was derived from
the dimensional reduction of the string motivated models, discussed without
details in section 2,
 and gives us,
in a more natural way, the dilaton-scalar matter coupling. The second
modification was motivated by the off-shell relations, which can be used in
a semiclassical version of the model.
The resulting theory is solved in
a way very similar to the CGHS model, showing in the  two studied cases
black-hole type solutions. We mention that the relation (24), present in both
the CGHS and the modified version presented here, it is important to make
the solutions in this simple way. In the new version, however, this happens
only when we consider the two modifications together.

 When there are two matter sources, as
given by (28) and (35) the situation is the one in which the solution is called
eternal black-hole, already discussed at the end of the section 3.
However, it is the first case, the one
 with the same source as CGHS model, that seems
to have more interesting features. As mentioned before, there is also
an horizon, defined by
the shift in the $x^-$ coordinate, but all the spacetime has a black-hole
behaviour, in contrast to the CGHS model, where only in the $x^+ > x^+_{0}$
region has this solution. The linear vacuum region $x^+<x^+_{0}$ is
important in the CGHS model because this condition  is used to determine
the Hawking  radiation. It seems  interesting to calculate this effect
in the picture given by this new solution. This work is in progress.

\bigskip
\bigskip
\bigskip
\pagebreak
\noindent
{\bf ACKNOWLEDGEMENTS}

The author is grateful to E.Abdalla for useful comments. This work was
supported by CNPQ (Brazil).

\bigskip
\bigskip
\bigskip
\noindent
{\bf REFERENCES}

\noindent [1] C.G. Callan, S.B. Giddings and J.A. Strominger,
Phys. Rev. D 45 (1992)
R1005; J.A. Strominger, in Les Houches Lectures on Black Holes, Les Houches
(1994)

\noindent [2] R. Mann, A. Shiekm and L. Tarasov, Nucl. Phys. B341 (1992) 134;
R. Jackiw, in Quantum Theory of Gravity, ed. S.Christensen (Adam
Hilger,Bristol,
1984), p.403; C. Teitelboim, ibid, p. 327.

\noindent [3] J. Russo, L. Susskind and L. Thorlacius,
Phys. Rev. D 45 (1992) 3444;
47 (1993) 533

\noindent [4] J. Maharana and J.H. Schwarz, Nucl. Phys. B390 (1993) 3;
J.Scherk and J.H.
Schwarz, Nucl. Phys. B153 (1979) 61; J. Maharana,
Phys. Rev. Lett. 75,2 (1995) 205.

\noindent [5] N.D. Birrel and P.C. Davies, in
Quantum Fields in Curved Spacetime
(Cambridge University Press, Cambridge, 1984)

\noindent [6] K. Fujikawa in Quantum Gravity and Cosmology, ed. H.Sato
and T.Inami (Singapore: World scientific); Phys. Rev. D 25 (1982) 2584.

\noindent [7] K. Fujikawa, U. Lindstrom, N.K. Rocek and P.van Nieuwenhuizen,
Phys. Rev. D 37 (1988) 391.

\end{document}